\documentclass[final]{IEEEtran}
\usepackage{amsthm,amssymb,graphicx,multirow,amsmath,color,amsfonts}
\usepackage[update,prepend]{epstopdf}
\usepackage[noadjust]{cite}
\usepackage[latin1]{inputenc}
\usepackage{tikz}
\usetikzlibrary{arrows,calc}		
\usepackage{bbm} 
\usepackage{pdfpages}
\usepackage{flushend}
\usepackage{tabulary}
\usepackage{multirow}
\usepackage{comment}

\def\nb0{{\mathbf{0}}}
\def\nb1{{\mathbf{1}}}

\newtheorem{lemma}{Lemma}

\newtheorem{theorem}{Theorem}

\newtheorem{remark}{Remark}

\def \barps {\bar{E_s}}
\def \grx {\Gamma_{rx}}

\allowdisplaybreaks 

\begin{document}
\graphicspath{{./Figures/}}
\title{
Novel LoS $\beta-\gamma$ THz Channel Unifying Molecular Re-radiation Manifestations 
}
\author{
 Anish Pradhan, J. Kartheek Devineni, Andreas F. Molisch, and Harpreet S. Dhillon
\thanks{A. Pradhan and H. S. Dhillon  are with Wireless@VT, Department of ECE, Virginia Tech, Blacksburg, VA, USA (email: \{pradhananish1, hdhillon\}@vt.edu). J. K. Devineni is with the Qualcomm Standards and Industry Organization, Qualcomm Technologies Inc., San Diego, CA 92121 USA. Email: jdevinen@qti.qualcomm.com. A. F. Molisch is with the Wireless Devices and Systems Group, Ming Hsieh Department of Electrical and Computer Engineering, University of Southern California, Los Angeles, CA, USA (email: molisch@usc.edu). This work was supported by U.S. National Science Foundation under Grants ECCS-2030215 and CNS-2107276.} 
\vspace{-8mm}
}

\maketitle

\begin{abstract}
This paper introduces a novel line-of-sight (LoS) $\beta-\gamma$ terahertz (THz) channel model that closely mirrors physical reality by considering radiation trapping. Our channel model provides an exhaustive modeling of the physical phenomena including the amount of re-radiation available at the receiver, parametrized by $\beta$, and the balance between scattering and noise contributions, parametrized by $\gamma$, respectively. Our findings indicate a nontrivial relationship between average limiting received signal-to-noise ratio (SNR) and distance emphasizing the significance of $\gamma$ in THz system design. We further propose new maximum likelihood (ML) thresholds for pulse amplitude modulation (PAM) and quadrature amplitude modulation (QAM) schemes, resulting in analytical symbol error rate (SER) expressions that account for different noise variances across constellation points. \textcolor{black}{The results confirm that the analytical SER closely matches the true simulated SER when using an optimal detector. As expected, under maximum molecular re-radiation, the true SER is shown to be lower than that produced by a suboptimal detector that assumes equal noise variances.}
\end{abstract}


\section{Introduction} \label{sec:intro}
Considered the final frontier of the radio spectrum, the terahertz (THz) band holds immense promise for delivering data rates in the range of multiple gigabits-per-second (Gbps) \cite{Tripathi2021}. In this vast and largely uncharted THz spectrum, physically faithful representation of molecular absorption and re-radiation has remained an open challenge. The way re-radiation is modeled critically affects its amount, and current models often overlook the subtleties involved in this phenomenon. Conventionally, these processes have been represented as \emph{sky-noise models} \cite{jornet_2012, AlouniTHz}, which significantly departs from their actual physical embodiment, namely the radiation trapping \cite{molisch1998radiation}. Conventional sky-noise models often treat all re-radiation as noise, a concept disputed by recent studies \cite{molabs,arxTHz}. This raises questions about the true nature of re-radiation in THz communication. On the one hand, re-radiation can correlate with the signal, acting as a useful scattering component that may enhance communication. On the other hand, it may act as a noise component, hindering the communication process. Current THz LoS channel models often fall short in capturing these characteristics of molecular re-radiation, neglecting both the underlying physics of its nature and the way it is manifested. Our main goal is to bridge this gap by developing an improved THz LoS channel model that truly captures these physical realities. In our model, we use radiation trapping to quantify the re-radiation amount and introduce a parameter $\gamma$ to represent the spectrum of its manifestation ranging from useful scattering all the way to Gaussian noise. The foundational groundwork laid out in this paper is expected to inspire future research in THz channel modeling, such as the estimation of the values of $\gamma$ in different operational situations.

\subsection{Related Work}  \label{sec:related}
In addition to the unavailability of physically accurate models for molecular re-radiation, this work is further motivated by the recent advancements in THz sources \cite{thzs1,thzs2,thzs3,probris} as well as interest in using this spectrum for new applications in nano-networks and indoor communication \cite{thznano,thznano2,thzindoor}. Two specific open research directions that influence our proposed $\beta$-$\gamma$ THz channel model are:  a) the physically accurate modeling of molecular re-radiation to characterize its amount, and b) a deeper examination of the inherent nature of re-radiation itself. 


The physical phenomenon that complicates modeling molecular re-radiation is called radiation trapping \cite{molisch1998radiation}. When a ray goes through atomic vapor, some atoms get excited through absorption of radiation, i.e., move to a higher energy state.  When they fall back to a lower energy state, the extra energy gets re-radiated as a resonance photon. Since its energy matches the transition energy, other atoms readily absorb it. This keeps happening until the re-radiation ultimately escapes the atomic vapor  (this is a rough description of the process, for a more exact formulation see \cite{molisch1998radiation}). Molecules also go through a similar absorption procedure \cite{discussion}. This re-radiation is modeled conventionally as sky-noise in THz communication literature \cite{jornet_2012}. It is a convenient way to interpret the re-radiated signal as noise. However, this model lacks validation \cite{AlouniTHz} and impacts the estimation of re-radiated power available at the receiver.

Further complicating the situation, research shows that when THz pulses pass through N$_{\text{2}}$O vapors, they produce coherent delayed pulses and cause molecules to orient towards the applied field in a Boltzmann distribution near absorption peaks \cite{Harde1, Harde2}. This hints at a correlation in the re-radiated pulse, a finding supported by several works \cite{jornet_2012, thznano2, molisch1998radiation}. Notably, \cite{molabs} recently proposed modeling the effective channel as a Rician channel with a Rician factor dependent on the absorption coefficient. Despite these advances, the question of whether re-radiated energy is wholly lost as noise or can be utilized as scattered signals remains under debate. As suggested by some authors, the actual effect can be thought of as a mixture of these two extremes \cite{discussion, arxTHz}. In light of these insights, our work aims to develop a channel model that estimates re-radiation using radiation trapping and controls the balance between scattering and noise contribution, incorporating the parameters $\beta$ and $\gamma$, respectively.

\subsection{Contributions}  \label{sec:contributions}
We introduce a novel LoS $\beta-\gamma$ THz channel model that focuses on the quantification of re-radiated power, moving away from the conventional \emph{sky-noise model}. This is done by factoring in radiation trapping \cite{molisch1998radiation} in conjunction with antenna theory, thereby providing a more accurate physical representation. We first include the parameter $\beta$ in the model, which approximates the fraction of the maximum re-radiated power that is captured at the Rx. Our results demonstrate that $\beta$ decreases with distance when the transmitter (Tx) and the receiver (Rx) are far apart, which is a consequence of the isotropic nature of re-radiation. We also demonstrate how the average limiting received SNR varies with distance, factoring in the influence of $\beta$. In addition, the parameter $\gamma$ is introduced to control the fraction of re-radiated power between scattering and noise contribution.

Furthermore, in THz communication, the amount of re-radiation is proportional to the transmit power of a symbol. Modulation schemes with symbols of different powers consequently result in varying noise variances associated with the constellation points, a detail often neglected. Recognizing this relationship, we offer new ML thresholds for PAM and QAM schemes, considering different noise variances for symbols of varying power. \textcolor{black}{Simulation results show that the analytical SER results based on the proposed ML thresholds (exact for PAM, and tight upper bound for QAM) match well with the simulated SER resulting from an optimal detector. As expected, this true SER is shown to be lower than the simulated SER when the symbols are detected through a suboptimal detector that assumes equal noise variance for all the symbols, particularly when the amount of molecular re-radiation is maximum.}
\section{Proposed Channel Model} \label{sec:ChanMod}
In traditional THz communication models, a fraction of the signal power, represented as $e^{-k(f)d}$, is assumed to reach the receiver through the LoS channel, while the remaining fraction, $1-e^{-k(f)d}$, is absorbed. The re-radiated signal is typically either modeled as additive Gaussian noise \cite{jornet_2012} or as a scattering component \cite{molabs}. Here, $k(f)$ represents the molecular absorption coefficient, $f$ denotes the operating frequency, and $d$ is the link distance. The polynomial model of the molecular absorption coefficient is defined by the absorption coefficient $y_i$ of $i$-th absorption line. It is dominated by the contribution of water vapor alone \cite{SimpleTHz}. The coefficient is given by $k(f) = \sum_i y_i(f, \mu) + g(f, \mu)$, where $\mu$ is the water vapor mixing ratio, and $g$ serves as an error-correcting polynomial. Further details, omitted for brevity, can be found in \cite{SimpleTHz}. While achieving these extreme manifestations may not be practical, they serve as limits for how much the re-radiation positively or negatively contributes to communication. Similar to the discussion in \cite{discussion}, we explore the possibility that the actual situation is a combination of these two extreme phenomena. 

Moreover, in both of these manifestations, existing literature implicitly assumes that all re-radiated power reaches the receiver. However, we propose a different approach incorporating the concept of ``radiation trapping.'' This allows us to mathematically compute the fraction $\beta$ of the maximum re-radiated power reaching the receiver, challenging the conventional assumption that sets $\beta=1$. This derivation is made under the assumption that \textcolor{black}{only single-absorption/reemission processes occur}, which is a reasonable approximation if non-radiative energy decay of excited molecules (quenching) occurs at a rate comparable to the natural decay rate, and/or the absorption coefficient is relatively low. A more exact formulation would require not only more complicated mathematical formulations but also a definition of the boundary of the volume in which the vapor occurs. Building on these assumptions, we present a relatively straightforward derivation of $\beta$ in the subsequent theorem utilizing an antenna-theoretic perspective.
\begin{theorem} \label{theo:beta}
    For a point-to-point THz link of distance $d$, beamwidth $\theta$ assuming a circular beam, and molecular absorption coefficient $k(f)$, the corresponding expression for $\beta$ is given by the following:\begin{align}
    &\beta=\frac{k(f) d^2}{2(1-e^{-k(f) d})} \notag \times \\& \int\limits_{\epsilon_1}^{d-\epsilon_2} \int \limits_{0}^{x\tan{\theta}}\!\!\frac{r\cos\vartheta e^{-k(f) \left(x+\sqrt{x^2+r^2}+\sqrt{(d-x)^2+r^2}\right)}}{((d-x)^2+r^2)(x^2+r^2)}\, {\rm d}r\, {\rm d}x. \label{eq:betaeq}
\end{align}
Here, the Rayleigh distances of the Tx and Rx are denoted as $\epsilon_1$ and $\epsilon_2$ respectively while $\cos{\vartheta}=\frac{d-x}{\sqrt{(d-x)^2+r^2}}$ denotes the projected portion of the effective aperture of the receiver. 
\end{theorem}
\begin{IEEEproof}
    See Appendix \ref{sec:Betaproof}.
\end{IEEEproof}
We further assume that a fraction $\gamma$ of the received re-radiated power contributes to the re-radiated scattering event and the remaining fraction $1-\gamma$ contributes to the noise variance. By incorporating these parameters, we can express a unified LoS THz channel as follows:
\begin{align}
{h}&= \sqrt{a} e^{j\theta} + \mathcal{CN}(0, \gamma\beta (1-a)),
\end{align}
where $ \quad \gamma \in [0,1),\quad \beta \in [0,1],$ and $a = e^{-k(f)d}$.  Alternatively, this channel can be expressed like a Rician channel where the Rician factor $K(f,d)$ of any link of distance $d$ is given by:
\begin{align}
    K(f,d)&=\frac{\text{Power of the LoS channel}}{\text{Power of the re-radiated component}}\notag\\&=\frac{e^{-k(f)d}}{\gamma\beta(1-e^{-k(f)d})}.\label{eq:RiceFactor}
\end{align}
Using this Rician factor, the LoS channel can be expressed in a Rician-like formulation as shown below:
\begin{align}
    &{h}=\begin{cases}
    \sqrt{\frac{K(f,d)}{K(f,d)+1}}\sigma_le^{j\theta}+\sqrt{\frac{1}{K(f,d)+1}}\mathcal{CN}(0,\sigma_l^2), &\mbox{if } \gamma > 0, \\
    \sigma_le^{j\theta}, &\mbox{if } \gamma = 0,
    \end{cases} \label{eq:Channel}
\end{align}
where $\sigma_l^2=e^{-k(f)d}+\gamma\beta(1-e^{-k(f)d})$ is the total power through the channel. Nonetheless, a comprehensive model necessitates characterizing the received noise, as the molecular re-radiation noise has an impact on the total noise level. The received noise follows this distribution:
\begin{align}
&n\sim\mathcal{CN}\left(0,\sigma^2+\barps\beta(1-\gamma)(1-e^{-k(f)d})\right), \label{eq:TotalNoiseVariance}
\end{align}
where $\sigma^2$ is the thermal noise and $\barps$ is the received power. As the channel can be expressed as a Rician channel, the probability distribution function (PDF) and cumulative distribution function (CDF) of the channel amplitude $r=|h|$ for $\gamma>0$  are known and given below ignoring  the frequency and distance dependence of the Rician factor for brevity \cite{molisch2023}:
\begin{align}
    &f_{R}(r)\!\!=\!\!\frac{2(\!K\!+\!1\!)xe^{\left(\!{-K-\frac{(K+1)r^2}{\sigma_l^2}}\!\right)}}{\sigma_l^2}\mathcal{I}_0\!\!\left(2r\!\sqrt{\frac{K(K+1)}{\sigma_l^2}}\right), \label{eq:chanPDF}\\
    &F_R(r)=1-\mathrm{Q}_1(\sqrt{2K},\frac{r\sqrt{2(K+1)}}{\sigma_l}), \label{eq:chanCDF}
\end{align}
where $\mathrm{Q}_1$ is the Marcum Q-function. Due to difficulties with modified Bessel functions $\mathcal{I}_0(\cdot)$ in numerical integrals, we use a normal approximation for $\sqrt{2K}\ggg1$ \cite{goodman2007speckle}. Thus, $r\sim\mathcal{N}\left(\sqrt{\frac{K}{K+1}}\sigma_l,\frac{1}{2(K+1)}\sigma_l^2\right)$ is used analytically.
\section{SNR Insights} \label{sec:SNRCapAnalysis}
This section provides SNR insights related to the molecular absorption coefficient and $\beta$ for this unified channel model in a simple point-to-point wireless link. For this link, the signal model is expressed as:
\begin{align}
y = h x+n,
\end{align}
where $y$ is the received signal, $x$ is the transmitted signal whose variance absorbs the effect of path-loss, $\mathrm{E}[|x|^2] =\bar{E}_s=\left(\frac{c}{4\pi f d}\right)^2 E_{\rm avg}$, and $E_{\mathrm{avg}}$ is the transmit power. Now, the instantaneous received SNR can be written as:
\begin{align}
    {\Gamma}=\frac{\barps r^2 }{\barps(1-\sigma_l^2)+\sigma^2}=\frac{\Gamma_{rx} }{\Gamma_{rx}\beta(1-\gamma)(1-a)+1}r^2, \label{eq:SNR}
\end{align}
where $\Gamma_{rx}=\frac{\barps}{\sigma^2}$ denotes the received signal-to-noise ratio (Rx SNR). Note that $\Gamma_{rx}$ only considers the thermal noise and is independent of the molecular absorption coefficient, and $\beta$.
\begin{lemma}
\label{lem:AsymSNR}
As $\grx\to\infty $, the average received SNR $\mathrm{E}[\Gamma]\to\frac{a+\gamma\beta(1-a)}{\beta(1-\gamma)(1-a)}$ when $\gamma<1$. In this operational regime, the limiting $\mathrm{E}[\Gamma]$ increases monotonically with respect to $\gamma$.
\end{lemma}
\begin{IEEEproof}
This can be readily found by taking limits on $\grx$ in equation \eqref{eq:SNR} and using $\mathrm{E}[r^2]=\sigma_l^2$ from the Rician distribution moments. Note that this holds even in the case of $\gamma=0$. The monotonic nature of the limiting average SNR can be trivially found by inspecting the respective derivative.
\end{IEEEproof}
\begin{remark}
\label{rem:ScatteringSNR}
When $\gamma = 1$, indicating that all absorbed signals are re-radiated as a useful scattering component, or when $\beta = 0$, signifying the absence of re-radiated absorption noise, the average received SNR approaches infinity as $\grx \to \infty$. However, in practical operating regimes, the average received SNR is bounded by an upper limit, given by $\frac{a+\gamma\beta(1-a)}{\beta(1-\gamma)(1-a)}$.
\end{remark}
\section{SER Analysis}
\label{sec:BERAnalysis}
While assessing the impact of molecular absorption and re-radiation on SNR for an LoS THz channel is straightforward, SER analysis is more complex. The complexity arises because each constellation point with different energy leads to different noise variances, as indicated by \eqref{eq:TotalNoiseVariance}. Meanwhile, the available SER results in the existing literature typically assume the same noise variance for all the constellation points. To fill in this gap, we provide a deeper SER analysis for both PAM and more complicated QAM schemes by considering different noise variances for constellation points with different energies.

\subsection{PAM}
PAM modulation is symmetric about the y-axis, so we only consider the constellation points on the positive x-axis. These points for PAM are represented as \((2i-1-m)\Delta\), with \(m\) ranging from 1 to \(M\). The value of \(\Delta\) is \(\sqrt{\frac{3 E_{\rm avg}}{M^2 -1}}\). The noise variance for each PAM symbol, \(\sigma_i^2\), is dependent on signal energy and described by \(\sigma^2 + (2i-1-M)^2 \Delta^2 \beta(1-\gamma) (1-a)\). Upon reception, the received symbol vector gets multiplied by the channel fading phase offset. The symbol vector for detection then becomes \(\tilde{y} = |h| x+\tilde{n}\), where the noise vector, \(\tilde{n}\), maintains its original statistics. By translating the origin to the midpoint of any two adjacent points, the threshold of the two points can be evaluated as follows:
\begin{align}
& \frac{\exp \left( \frac{-(x+|h|\Delta)^2}{\sigma_i^2} \right)}{\sqrt{\pi \sigma_i^2}} \gtrless_{i+1}^{i} \frac{\exp \left( \frac{-(x-|h|\Delta)^2}{\sigma_{i+1}^2} \right)}{\sqrt{\pi \sigma_{i+1}^2}} \nonumber\\
&\implies {x}_{i}^{i+1} \gtrless_{i+1}^{i} - \left( \frac{\sigma_{i+1}^2 + \sigma_i^2}{\sigma_{i+1}^2 - \sigma_i^2} \right) |h| \Delta \pm \nonumber \\ &\sqrt{\frac{\sigma_{i+1}^2 \sigma_i^2}{2(\sigma_{i+1}^2 - \sigma_i^2)} \ln \left(\frac{\sigma_{i+1}^2}{\sigma_i^2}\right) + |h|^2 \Delta^2 \left( \left( \frac{\sigma_{i+1}^2 + \sigma_i^2}{\sigma_{i+1}^2 - \sigma_i^2} \right)^2 \!\!\!\!-\!1\right) }.
\end{align}
The ML threshold from this decision rule can be written as
\begin{align}
&T^i_{i+1} = - \left( \frac{\sigma_{i+1}^2 + \sigma_i^2}{\sigma_{i+1}^2 - \sigma_i^2} \right) |h| \Delta+ \nonumber\\
&\sqrt{\!\frac{\sigma_{i+1}^2 \sigma_i^2}{2(\sigma_{i+1}^2 - \sigma_i^2)} \ln \left(\frac{\sigma_{i+1}^2}{\sigma_i^2}\right) + |h|^2 \Delta^2 \left( \left( \frac{\sigma_{i+1}^2 + \sigma_i^2}{\sigma_{i+1}^2 - \sigma_i^2} \right)^2\!\!\!\!-\!1\right)}.
\end{align}

The symbol error rate (SER) for the PAM modulation in this scenario can be evaluated as
\begin{align}
&P_{\rm PAM} = \frac{2}{M} \Bigg(Q\left(\frac{|h|\Delta}{\sigma_{\frac{M}{2}}/\sqrt{2}} \right)+ \nonumber\\
& \sum_{i=\frac{M}{2}+1 }^{M-1} \left[ Q \left( \frac{T^i_{i+1} + |h|\Delta}{\sigma_{i}/\sqrt{2}}\right) + Q \left( \frac{|h|\Delta - T^i_{i+1}}{\sigma_{i+1}/\sqrt{2}}\right)\right]\Bigg).
\end{align}

\subsection{QAM}
We consider a square QAM constellation $(M=4, 16, 64, 256)$ for the analysis but it can be easily extended to  rectangular constellations with appropriate changes. Due to symmetry, finding the SER of the constellation points in one quadrant is sufficient. One quadrant of the square M-QAM constellation can be divided into $(\sqrt{M}/2-1)^2$ middle points, $\sqrt{M}/2-1$ upper and lower side points, and $1$ corner point. These points can be indexed by $\left((2i-1)\Delta,(2j-1)\Delta\right)$, where $\Delta = \sqrt{\frac{3 E_{\rm avg}}{2(M -1)}}$. The noise variance of a QAM symbol is dependent on the signal energy and is given by $\sigma_{i,j}^2 = \sigma^2 + \left((2i-1)^2 + (2j-1)^2\right) \Delta^2 \beta(1-\gamma) (1-a)$. Now, the ML threshold $T(i,j,i',j')$ between $\left((2i-1)\Delta,(2j-1)\Delta\right)$ and $\left((2i'-1)\Delta,(2j'-1)\Delta\right)$ can be expressed as:
\begin{align}
    T_{i,j}^{i',j'}=\frac{-B \pm \sqrt{B^2-4AC}}{2A}, \label{eq:ThresholdQam}
\end{align}
where, $A=\sigma^2_{i,j}-\sigma^2_{i',j'}$, $B=2(p_0\sigma^2_{i',j'}-p_1\sigma^2_{i,j})$ and $C=(p_1^2\sigma^2_{i,j}-p_0^2\sigma^2_{i',j'})-\log\frac{\sigma_{i,j}}{\sigma_{i',j'}}\sigma^2_{i,j}\sigma^2_{i',j'}$. For horizontal thresholds, $p_0=(2i-1)|h|\Delta$ and $p_1=(2i'-1)|h|\Delta$. For vertical thresholds, $p_0=(2j-1)|h|\Delta$ and $p_1=(2j'-1)|h|\Delta$. Now, the union bound of the SER considering only nearest neighbors for the QAM scheme can be evaluated as:
\begin{align}
    P_{\rm QAM}\leq P_{\rm QAM,union}=
    \frac{4}{M}(P_m+P_s+P_c),
\end{align}
where, $P_m$ is the SER of the constellation points with $4$ adjacent points, $P_s$ is the SER of the constellation points with $3$ adjacent points and $P_c$ is the SER of the cornermost constellation point elaborated in \eqref{eq:Pm}, \eqref{eq:Ps}, and \eqref{eq:Pc}.
\begin{figure*}
\begin{align}
    P_m\!=\!&\sum\limits_{j=1}^{\frac{\sqrt{M}}{2}-1}\!\sum\limits_{i=1}^{\frac{\sqrt{M}}{2}-1}\!\left[1\!-\!\left(\mathrm{Q}\left(\frac{T_{i,j}^{i+1,j}\!-\!p_0(i)}{\sigma_{i,j}/\sqrt{2}}\right)\!-\!\mathrm{Q}\left(\frac{T_{i,j}^{i-1,j}\!-\!p_0(i)}{\sigma_{i,j}/\sqrt{2}}\right)\right)\!\!\!\left(\mathrm{Q}\left(\frac{T_{i,j}^{i,j+1}\!-\!p_0(j)}{\sigma_{i,j}/\sqrt{2}}\right)\!-\!\mathrm{Q}\left(\frac{T_{i,j}^{i,j-1}\!-\!p_0(j)}{\sigma_{i,j}/\sqrt{2}}\right)\right)\right], \label{eq:Pm} \\
    P_s=&2\sum\limits_{i=1}^{\frac{\sqrt{M}}{2}-1}\left[1-\left(\mathrm{Q}\left(\frac{T_{i,\frac{\sqrt{M}}{2}}^{i+1,\frac{\sqrt{M}}{2}}-p_0(i)}{\sigma_{i,\frac{\sqrt{M}}{2}}/\sqrt{2}}\right)-\mathrm{Q}\left(\frac{T_{i,\frac{\sqrt{M}}{2}}^{i-1,\frac{\sqrt{M}}{2}}-p_0(i)}{\sigma_{i,\frac{\sqrt{M}}{2}}/\sqrt{2}}\right)\right)\left(1-\mathrm{Q}\left(\frac{T_{i,\frac{\sqrt{M}}{2}}^{i,\frac{\sqrt{M}}{2}-1}-p_0(\frac{\sqrt{M}}{2})}{\sigma_{i,\frac{\sqrt{M}}{2}}/\sqrt{2}}\right)\right)\right], \label{eq:Ps}\\
    P_c=&1-\left(1-\mathrm{Q}\left(\frac{T_{\frac{\sqrt{M}}{2},\frac{\sqrt{M}}{2}}^{\frac{\sqrt{M}}{2}-1,\frac{\sqrt{M}}{2}}-p_0(\frac{\sqrt{M}}{2})}{\sigma_{\frac{\sqrt{M}}{2},\frac{\sqrt{M}}{2}}/\sqrt{2}}\right)\right)\left(1-\mathrm{Q}\left(\frac{T_{\frac{\sqrt{M}}{2},\frac{\sqrt{M}}{2}}^{\frac{\sqrt{M}}{2},\frac{\sqrt{M}}{2}-1}-p_0(\frac{\sqrt{M}}{2})}{\sigma_{\frac{\sqrt{M}}{2},\frac{\sqrt{M}}{2}}/\sqrt{2}}\right)\right). \label{eq:Pc}
\end{align}
\hrule
\vspace{-4.5mm}
\end{figure*}
\section{Numerical Results} \label{sec:NumResults}
\begin{figure*}
 \centering
\begin{minipage}[b]{0.24\linewidth}
 \centering
    \includegraphics[width=\textwidth]{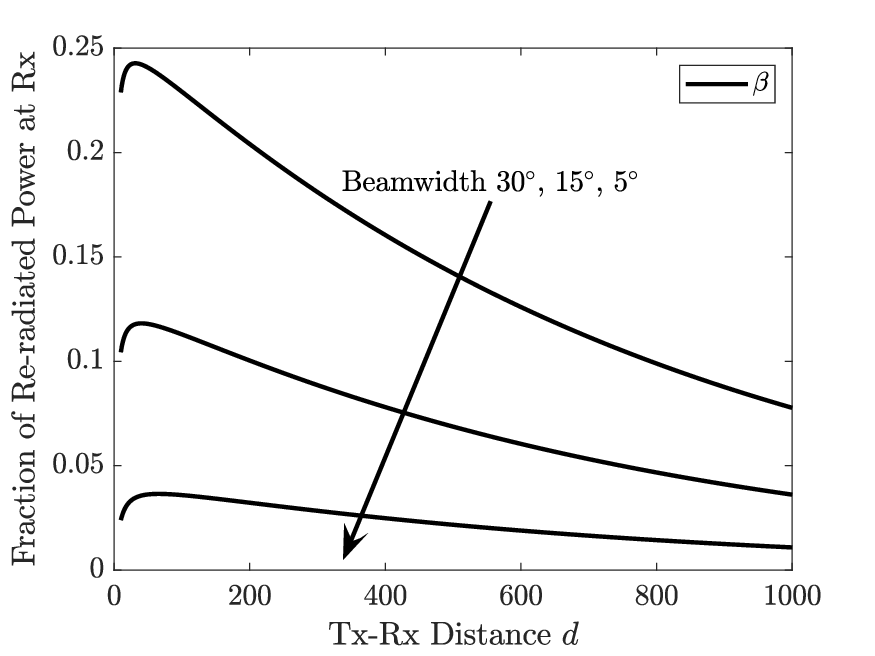}
    \caption{Derived $\beta.$}
    \label{fig:Rx}
\end{minipage}
\begin{minipage}[b]{0.24\linewidth}
 \centering
    \includegraphics[width=\textwidth]{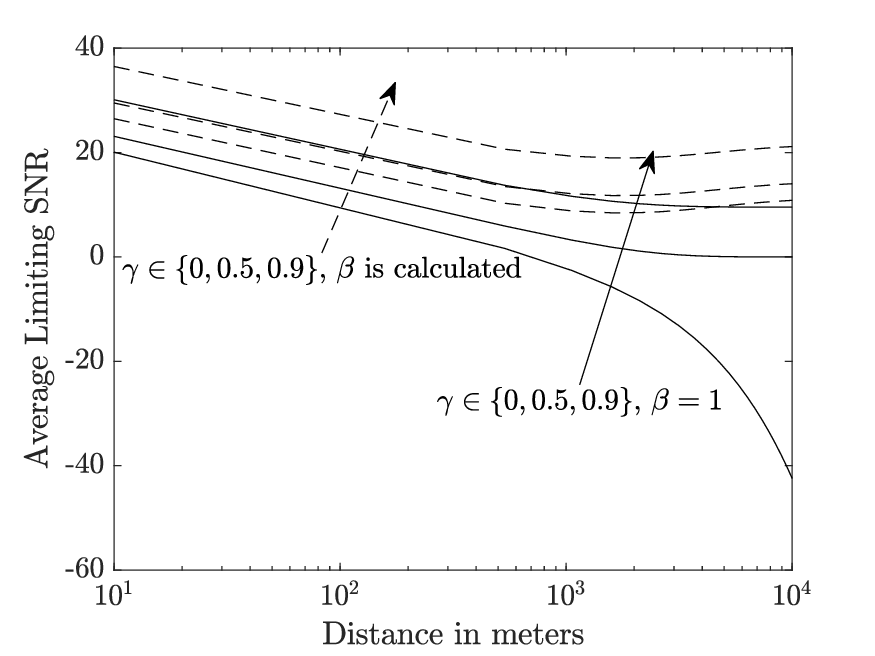}
    \caption{Average Limiting SNR.}
    \label{fig:LSNR}
\end{minipage}
\begin{minipage}[b]{0.24\linewidth}
\centering
    \includegraphics[width=\textwidth]{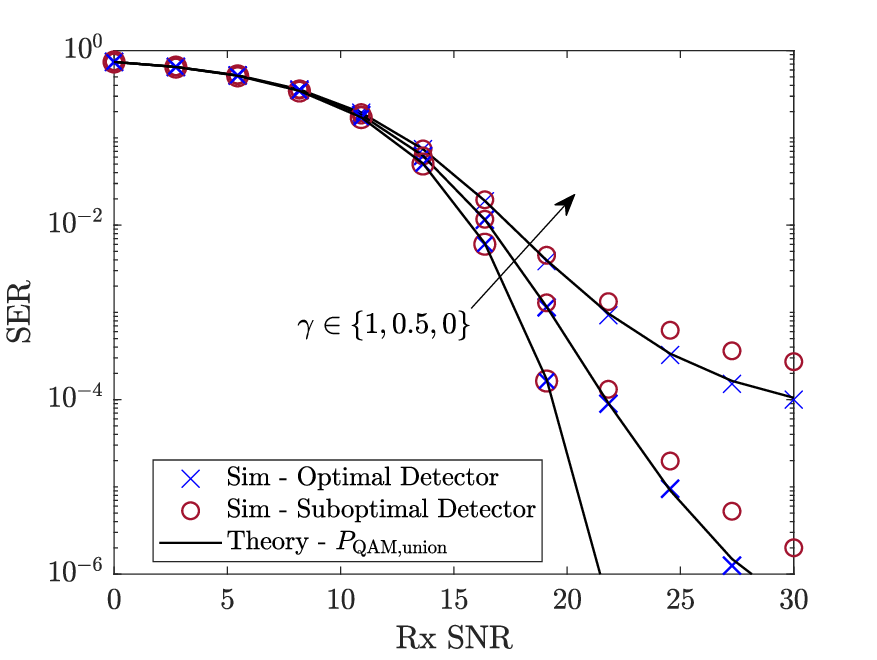}
    \caption{SER of 16-QAM when $\beta=1$.}
    \label{fig:QAMbb1}
\end{minipage}
\begin{minipage}[b]{0.24\linewidth}
\centering
\includegraphics[width=\textwidth]{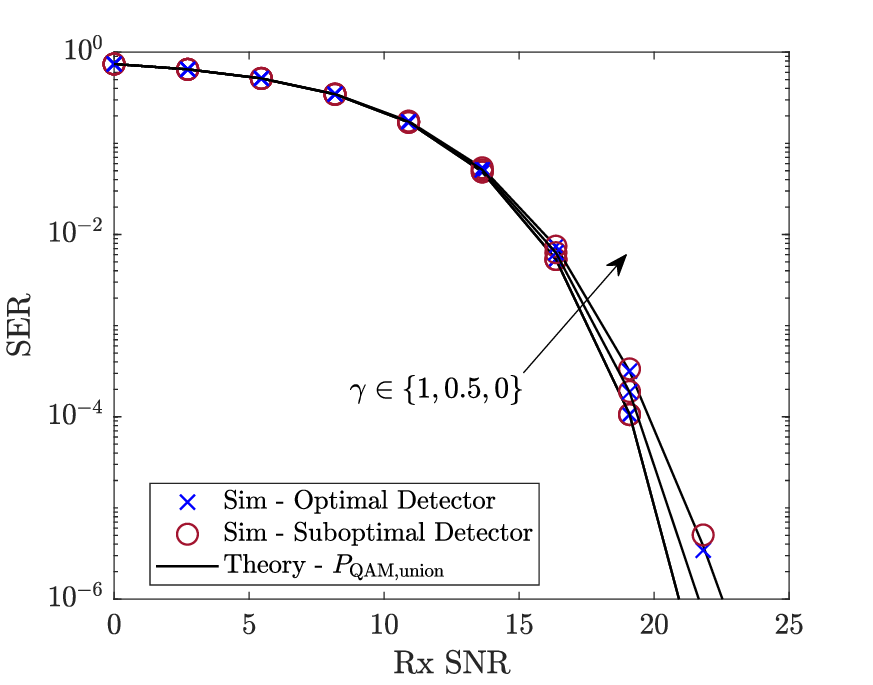}
    \caption{SER of 16-QAM when pracical $\beta$ is used.}
    \label{fig:QAMbbc}
    \end{minipage}
    \vspace{-6mm}
\end{figure*}
In this section, we explore the effects of varying $\gamma$ and $\beta$ parameters on SNR and SER for a point-to-point THz link, grounded in the practical aspects of THz link budget analysis as detailed in \cite{THzlink}. We simulate a communication link at $300$ GHz over a $10$ meter distance under $50\%$ relative humidity, $27^o$C temperature, and $1$ atm pressure. The absorption coefficient $k(f)$ is calculated according to \cite{SimpleTHz}. To ensure meaningful insights from the simulation, we use a realistic $\Gamma_{rx}$ range of $0\,{\rm dB}$ to $30\,{\rm dB}$, factoring in all transceiver architecture effects, including gain \cite{THzlink}. For the calculation of $\beta$ through equation \eqref{eq:betaeq}, the Rayleigh distances are taken to be $\epsilon_1=0.64$ m and $\epsilon_2=0.51$ m by assuming practical rectangular arrays.

Fig. \ref{fig:Rx} presents $\beta$ as a function of the Tx-Rx distance. We observe that $\beta$ increases briefly with increasing distance and then decreases again. Our intuition is that, for smaller distances, as the distance increases, the amount of re-radiated power increases due to the increase in absorption, and most of the power is available at the Rx. However, for larger distances, as the distance increases, the isotropic nature of the re-radiation dominates and $\beta$ decreases. 

The average limiting received SNR from Lemma \ref{lem:AsymSNR} is plotted in Fig. \ref{fig:LSNR} as a function of the link distance. We can observe that when $\beta=1$, the average limiting received SNR ${\rm E}[\Gamma]$ decreases rapidly with distance when $\gamma=0$ as all of the re-radiated power appears as noise. In this case, the numerator of ${\rm E}[\Gamma]=\frac{a}{1-a}$ is the transmittance that decreases with distance and the denominator is the noise amount that increases with distance. However, when $\gamma\neq 0$, ${\rm E}[\Gamma]\to \frac{\gamma}{1-\gamma}$ as $d \to \infty$. This can also be observed in the plot. \textcolor{black}{Next, when a realistic value of $\beta$ is calculated from Theorem \ref{theo:beta}}, we observe that ${\rm E}[\Gamma]$ reaches a minimum and then saturates. This becomes clear when examining the $\gamma=0$ case. In this case, ${\rm E}[\Gamma]=\frac{a}{\beta(1-a)}$ and we plot $10\log_{10}{\rm E}[\Gamma]=10\log_{10} a - 10\log_{10} \beta(1-a)$. The slope of the first term is a constant $-\frac{10}{\ln{10}}k(f)$ while we have observed that the second term denotes the amount of the re-radiated noise that first increases slightly with distance and then decreases rapidly. The behavior follows the rationale from the prior discussion. The slope of this term transitions from positive to negative. When it matches the first term's constant slope, the \({\rm E}[\Gamma]\) reaches its minimum.

\textcolor{black}{In Figures \ref{fig:QAMbb1} and \ref{fig:QAMbbc}, we compare the SER against Rx SNR for two detectors: an optimal one considering unequal noise variances, and a suboptimal one assuming equal noise variances across constellation points with different energy. As shown in Fig. \ref{fig:QAMbb1}, for $\beta=1$, the optimal detector yields a lower SER at higher Rx SNR values. The large variation of SER also highlights the significant impact of $\gamma$, especially in indoor scenarios where $\beta$ approaches unity. As $d=10$ m in our simulation, the derived $\beta$ value from Theorem \ref{theo:beta} is only $0.23$, making both detectors comparably efficient due to minimal re-radiation noise in Fig. \ref{fig:QAMbbc}.}
\section{Conclusions}
In this paper, we proposed a novel LoS $\beta-\gamma$ THz channel model that better reflects physical reality. We parameterize the re-radiation process using $\beta$ and $\gamma$, which characterize the amount of re-radiation, and determine its manifestation as noise and scattering components, respectively. The simulation results demonstrate a nontrivial relationship between average limiting received SNR and distance when a realistic value of $\beta$ is used and underscore the importance of $\gamma$ in THz system design. \textcolor{black}{Moreover, by proposing new ML thresholds for the optimal detectors of PAM and QAM schemes, which account for different noise variances among symbols of varying power, we obtain analytical SER expressions that match well with the simulation results. As expected, this SER outperforms the SER resulting from the suboptimal detector, especially when the amount of molecular re-radiation is at the theoretical maximum.} Future research could focus on extending this model to non-line-of-sight (NLoS) scenarios and measurement campaigns for estimating $\gamma$ in a variety of situations.
\appendix
\subsection{Proof of Theorem \ref{theo:beta}} \label{sec:Betaproof}
\begin{figure}
    \centering
    \includegraphics[trim={7.4cm 6cm 9.9cm 5.6cm},clip,width=0.6\columnwidth]{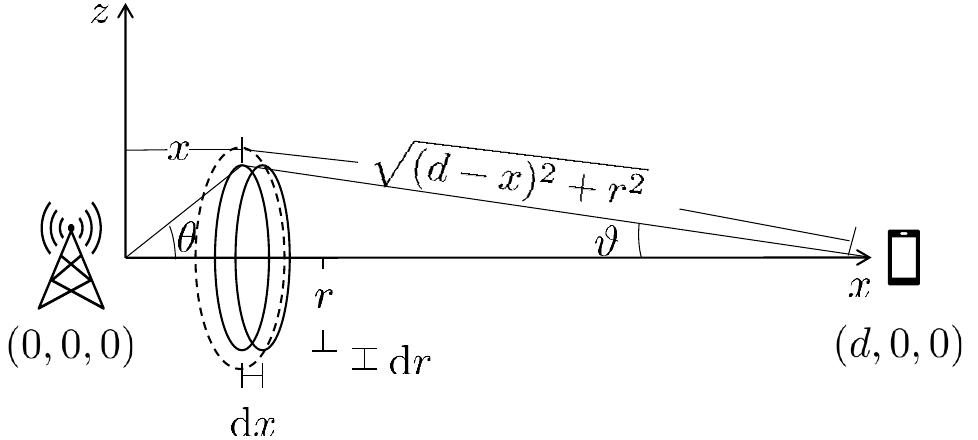}
    \caption{Power absorbed in a differential strip of width ${\rm d}x$.}
    \label{fig:Betacalc}
\end{figure}
We begin by assuming that the power is uniform within a cone whose opening angle corresponds to the antenna gain (i.e., a gain pattern that is a step function with respect to the angle with the x-axis), as depicted in Fig. \ref{fig:Betacalc}, the power absorbed and re-radiated in the annular strip of length ${\rm d}x$ is:
\begin{align}
    P_{x}&=\frac{P_{Tx}e^{-k(f)\sqrt{x^2+r^2}}}{4\pi (x^2+r^2)}\, G_{Tx}\, (e^{-k(f) x}\!-\!e^{-k(f) (x+{\rm d}x)})\, A_{strip} \notag \\ & \approx \frac{P_{Tx}e^{-k(f)\sqrt{x^2+r^2}}}{4\pi (x^2+r^2)}\, G_{Tx}\, k(f) \, e^{-k(f) x}\,A_{strip}\,{\rm d}x.
\end{align}
Assuming a narrow beamwidth, the received power at the Rx from a differential element is expressed as:
\begin{align}
    P_{Rx} \approx & \notag \frac{P_{Tx}e^{-k(f)\sqrt{x^2+r^2}}}{4\pi (x^2+r^2)}\, k(f) \, e^{-k(f) x}\,G_{Tx}\,A_{strip}\, A_{Rx}\, \times \\ &\cos\vartheta\, \frac{e^{-k(f) \sqrt{(d-x)^2+r^2}}}{4\pi((d-x)^2+r^2)}\,{\rm d}x,
\end{align}
where $\cos{\vartheta}=\frac{d-x}{\sqrt{(d-x)^2+r^2}}$ denotes the projected portion of $A_{Rx}$, and the area of the annular strip is $A_{strip}=2\pi r {\rm d}r$.
Assuming no secondary reabsorptions occur and neglecting the electromagnetic interactions within the Rayleigh distances of $\epsilon_1$ and $\epsilon_2$, total re-radiated power arriving at Rx is:
\begin{align}
    &P_{\rm diffused} = \notag \frac{k(f) P_{Tx}\, G_{Tx} A_{Rx}\,}{8\pi}\,\times \\&\int\limits_{\epsilon_1}^{d-\epsilon_2} \int \limits_{0}^{x\tan{\theta}}\!\!\!\!\!\frac{r\cos\vartheta e^{-k(f) \left(x+\sqrt{x^2+r^2}+\sqrt{(d-x)^2+r^2}\right)}}{((d-x)^2+r^2)(x^2+r^2)}\, {\rm d}r\, {\rm d}x. \label{eq:Pt}
\end{align}
Traditionally, the maximum re-radiated power at Rx is \cite{coint}:
\begin{align}
    P_{\rm diffused, max}=\frac{P_{Tx}}{4\pi d^2} G_{Tx} A_{Rx}(1-e^{-k(f) d}).
\end{align}
The fraction $\beta$ that is available at the Rx is expressed below:
\begin{align}
    &\beta=\notag\frac{P_{\rm diffused}}{P_{\rm diffused, max}}=\frac{k(f) d^2}{2(1-e^{-k(f) d})}\times \\&\int\limits_{\epsilon_1}^{d-\epsilon_2} \int \limits_{0}^{x\tan{\theta}}\!\!\!\!\!\frac{r\cos\vartheta e^{-k(f) \left(x+\sqrt{x^2+r^2}+\sqrt{(d-x)^2+r^2}\right)}}{((d-x)^2+r^2)(x^2+r^2)}\, {\rm d}r\, {\rm d}x. \label{eq:betaeqinproof}
\end{align}
This concludes the proof.
\IEEEQED
\bibliographystyle{IEEEtran}
\bibliography{hokie}
\end{document}